# A Computational Proof of the Highest-Scoring Boggle Board

Dan Vanderkam

2025-06-08

**Abstract**  Finding all the words on a Boggle board is a classic computer programming problem. With a fast Boggle solver, local optimization techniques such as hillclimbing and simulated annealing can be used to find particularly high-scoring boards. The sheer number of possible Boggle boards has historically prevented an exhaustive search for the global optimum board. We apply Branch and Bound and a decision diagram-like data structure to perform the first such search. We find that the highest-scoring boards found via hillclimbing are, in fact, the global optima.

## Introduction

Boggle is a word search game invented in 1972 by Allan D. Turoff and currently sold by Hasbro. Competitors shake up 16 dice to form a 4x4 grid of letters. The goal is to find as many words as possible in three minutes. Letters can be connected up, down, left, right and diagonally, and words need not be arranged in a straight line. The same die cannot be used twice in a word. Longer words count for more points (3/4 letters=1 point, 5=2, 6=3, 7=5, 8+=11 points).

Finding all the words on a Boggle board has become a classic computer programming problem.[1], [2] It is often assigned in classes[3], used as an interview question and, more recently, given as a task to LLMs.[4] With a fast Boggle solver, it's natural to search for particularly high-scoring boards. Typically this is done via hillclimbing, simulated annealing or genetic algorithms. Searches of this kind date back to at least 1982.[5] While these searches do produce high-scoring boards, they cannot make definitive statements about whether these are *the* highest-scoring boards.

This paper takes a different approach. By using Branch and Bound, a decision tree-like data structure tailor-made for Boggle, and a large amount of compute, we're able to establish for the first time that the best Boggle boards found via hillclimbing are, in fact, the global optima.

| P | E | R | S |
|---|---|---|---|
| L | A | T | G |
| S | I | N | E |
| T | E | R | S |

Table 1: The highest-scoring Boggle board for the ENABLE2K dictionary, with 1,045 words and 3,625 points. The longest word is "replastering."

## Terminology and Conventions

The words that can be found on a Boggle board are determined by the letters on the board and by the choice of dictionary. In this paper, we'll use the ENABLE2K word list, which was developed by Alan Beale in 1997 and 2000. This word list contains 173,528 entries. Many wordlists are designed for Scrabble, and do not contain words longer than 15 letters. ENABLE2K has no limit on word length.

We adopt the following terminology and conventions:

- $B$ refers to a Boggle board. To refer to a specific Boggle board, we write out the letters of the board in row-major order. (This distinction is only important for non-square dimensions such as 2x3 and 3x4 Boggle, which lack rotational symmetry.)
- Because one of the Boggle dice contains a "Qu" (two letters), we adopt the convention that `q` indicates a Qu cell. So `rpqaselinifcoita` refers to the board in Table 23.

- Boggle dice use uppercase letters (except for Qu), but we typically use lowercase. No meaningful distinction is drawn between uppercase and lowercase in this paper.
- The cells on an $M$x$N$ board are numbered $0...MN-1$ in row-major order, as shown in Table 2. We refer to the letter on cell $i$ of board $B$ as $B_i$.
- Words($B$) is the set of all words that can be found on the board $B$.
- $S(B)$ is the sum of the point value of these words,

$$S(B) = \sum_{w \in \text{Words}(B)} \text{SCORES}[\text{Len}(w)]$$

We refer to this as the score of the board.

| 0 | 1 | 2 | 3 |
| 4 | 5 | 6 | 7 |
| 8 | 9 | 10 | 11 |
| 12 | 13 | 14 | 15 |

Table 2: Numbering of cells on a 4x4 board.

## Heuristics to find high-scoring boards

Finding all the words on a Boggle board is a popular programming puzzle. This is done by performing a depth-first search over the board's adjacency graph starting at each cell. The key to making this efficient is to prune out prefixes such as "bn" that don't begin any words in the dictionary. This is typically done using a Trie (Prefix Tree) data structure.

```
# Listing 0: Scoring a Boggle Board
def score(bd: str, trie: Trie) -> int:
  used = {}

  def step(idx: int, node: Trie) -> int:
    score = 0
    used[idx] = True
    if node.has_child(bd[idx]):
      n = node.child(bd[idx])
      if n.is_word() and not n.is_visited():
        score += SCORES[n.length()]
        n.set_visited()
      for n_idx in NEIGHBORS[idx]:
        if not used.get(n_idx):
          score += step(n_idx, n)
    used[idx] = False
    return score

  return sum(
    step(i, trie) for i in range(m * n)
  )
```

Nodes in the Trie are marked as we find words to avoid double-counting. With some care, it is possible to find the score of individual boards extremely rapidly on modern hardware. For example, the author's M2 MacBook is able to score random 4x4 Boggle boards at a rate of around 200,000 boards per second.

This speed can be used to attack a new problem: finding high-scoring boards. This is typically done via local search heuristics such as hillclimbing[6], simulated annealing[7], or genetic algorithms.[8] A particularly effective approach for Boggle is to iteratively explore around a pool of high-scoring boards, as shown in Listing 1.

```
# Listing 1: Hillclimbing Algorithm
μ = 500  # pool size
Pool = μ random Boggle Boards
Repeat until convergence:
  Next = Pool + Boards within edit distance 1
  Pool = μ highest-scoring boards in Next
```

This is similar to a $(\mu + \lambda)$-Evolutionary Strategy,[9] the differences being that we evaluate all neighbors (rather than a random sample) and terminate after convergence. Given the initial pool of boards, this converges to a deterministic result.

We take an "edit" to mean changing one letter or swapping two letters. With a pool size of μ=500, this usually (95/100 times) converges with the highest-scoring board as `perslatgsineters` (Table 1), which contains 1,045 words and scores 3,625 points using ENABLE2K.

This consistent result makes one wonder whether this board is, in fact, the global optimum. The principal result of this paper is to show that it is.

## Proof of Maximality

The most straightforward way to prove that a board is the highest-scoring is by exhaustive search. Unfortunately, the combinatorial explosion of possible boards renders this infeasible for all but the smallest sizes:

| Dims | Num Boards | Rate | Time |
| --- | --- | --- | --- |
| 3x3 | $26^9/8 \approx 6.8\text{e}11$ | 600k bd/s | ~12 days |
| 3x4 | $26^{12}/4 \approx 2.4\text{e}16$ | 400k bd/s | ~2000 years |
| 4x4 | $26^{16}/8 \approx 5.5\text{e}21$ | 200k bd/s | ~900M years |

One objection is that not all $26^{16}$ combinations of letters can be rolled with the standard 16 Boggle dice. Determining whether a particular letter combination can be rolled is a Set-Cover problem, which is NP-Complete. A greedy approach works well for this small problem, however, and we can estimate that approximately 1 in 12 combinations of letters can be rolled. While this reduces the search space, it's not enough to make exhaustive search feasible, and this will not prove to be a useful constraint. In practice, all high-scoring boards can be rolled in many ways with the Boggle dice.

Two other observations suggest a path towards a solution:

1. Similar boards have similar sets of words on them. Finding the score of a board and that of its neighbors involves large amounts of duplicated work.
2. Most boards have scores that are significantly lower than the best one. The average score of a random board is about 42 points, nearly 100x less than 3,625. (The average score of a board rolled with Boggle dice is closer to 120 points.)

The first observation suggests that we should group boards together to reduce repeated work. The second indicates that we have considerable "wiggle room" to upper bound the score rather than calculate it precisely.

**Branch and Bound**
Rather than exhaustive search, we use Branch and Bound to find the globally optimal board. Branch and Bound is an algorithm design paradigm dating back to the 1960s that narrows in on the global optimum by recursively subdividing the search space.[10]

To apply it, we need to define two operations on sets of Boggle boards:

- bound($S$): an upper bound on the score of any board in a set of boards S
- branch($S$): a way to split the set S into smaller sets $S_1, S_2, ..., S_m$.

With these operations in place, the Branch and Bound algorithm to find all boards with a score greater than $S_\text{high}$ is given in Listing 2:

```
# Listing 2 - Branch and Bound Algorithm
Queue <- {Universal Set of MxN Boggle Boards}
while Queue is non-empty:
  S <- Pop(Queue)
  if |S| == 1:
    S is a candidate solution;
    Calculate Score(S) to confirm
  else if Bound(S) < S_high:
    S cannot contain a high-scoring board.
  else:
    for S_i in branch(S):
      Queue <- S_i
```

The appeal of this approach is that, when bound($S$) is low, we can discard the entire set $S$ without having to evaluate every board in it.

We need to define the branch and bound operations, but first we'll consider sets of Boggle boards.

**Board classes and the branch operation**
Rather than allowing arbitrary sets of Boggle boards, we restrict ourselves to "board classes." These require each cell of a board in the set to come from a particular set of letters:

$$C(L_1, L_2, ..., L_{mn}) = \{B \mid B_i \in L_i \forall i\}$$

For example, here's a 3x3 board class where each cell can be one of two letters:

| {A, B} | {G, H} | {M, N} |
|---|---|---|
| {C, D} | {I, J} | {O, P} |
| {E, F} | {K, L} | {Qu, R} |

Table 4: A 3x3 board class containing 512 boards

This board class contains $2^9$ = 512 possible boards. Here are a few of them:

| A | G | M |
|---|---|---|
| D | I | P |
| E | L | R |

85 points

| B | H | M |
|---|---|---|
| C | I | P |
| E | L | R |

62 points

| B | H | M |
|---|---|---|
| D | J | P |
| F | K | Qu |

0 points

Analogous to the $B_i$ notation for boards, we can indicate the possible letters on a cell in a board class as $C_i$. On this board class, for example, $C_0$ = {"A", "B"}.

We can carve the 26 letters of the alphabet up into distinct "buckets" to reduce the combinatorial explosion of possible boards into a much smaller number of board classes.

The following partitions of the alphabet were found via a heuristic search:

| N | Letter Buckets |
|---|---|
| 2 | {aeiosuy, bcdfghjklmnpqrtvwxz} |
| 3 | {aeijou, bcdfgmpqvwxz, hklnrsty} |
| 4 | {aeiou, bdfgjqvwxz, lnrsy, chkmpt} |

Using three buckets with 4x4 Boggle, for example, we have only $3^{16}/8 \approx 5.4 \times 10^6$ board classes to consider, a dramatic reduction from the $5.5 \times 10^{21}$ individual boards. This will be in vain if operations on board classes are proportionally slower, but fortunately, this will not prove to be the case.

To "branch" from a board class, we split one of its cells into the possible letters. For example, starting with this board class containing 5,062,500 boards:

| lnrsy | chkmpt | lnrsy |
|---|---|---|
| aeiou | aeiou | aeiou |
| chkmpt | lnrsy | bdfgjqvwxz |

Table 6: A 3x3 board class containing 5M boards.

we might split the center cell to get five board classes containing 1,012,500 boards each:

| lnrsy | chkmpt | lnrsy |
|---|---|---|
| aeiou | a | aeiou |
| chkmpt | lnrsy | bdfgjqvwxz |

| lnrsy | chkmpt | lnrsy |
|---|---|---|
| aeiou | e | aeiou |
| chkmpt | lnrsy | bdfgjqvwxz |

...

| lnrsy | chkmpt | lnrsy |
|---|---|---|
| aeiou | u | aeiou |
| chkmpt | lnrsy | bdfgjqvwxz |

Since the center and edge cells have the greatest connectivity, we split these first before splitting the corners.

Board classes still have all the same symmetries as a Boggle board. This allows us to only consider "canonically-oriented" board classes for a roughly 8x reduction in the search space.

### Sum/Choice trees

Our goal is to calculate an upper bound on a board class, and to implement a "branch" operation that can be performed efficiently.

It is possible to compute an upper bound on a board class directly using a DFS similar to score. And while this uses minimal RAM, it is not conducive to efficient branching. Instead, we'll develop a tree structure that's well-suited to this problem.

Our tree structure consists of alternating layers of two types of nodes:

```
Node := SumNode | ChoiceNode

ChoiceNode:
  cell: int
  children: {letter -> SumNode}

SumNode:
  points: int
  children: ChoiceNode[]
```

SumNode.points is the points on an individual node, not a bound for the entire subtree.

These nodes model the two types of branching for Boggle with board classes. A ChoiceNode models a choice of letter on a cell in a board class. A SumNode models how we can move in different directions from a cell, or start from any cell on the board. A path to a word in a board class is modeled as a path through the Sum/Choice tree.

We need to define a few basic operations on these trees. We can define a path as a sequence of cells and letters on those cells:

```
Path = list[(int, char)]  # (cell, letter)
```

Then we can define add_word:

```python
# Listing 3: add_word to sum/choice tree
def add_word(
  node: SumNode, path: Path, points: int
):
  if len(path) == 0:
    node.points += points
    return node

  cell, letter = path[0]
  choice_n = find(
    node.children, lambda c: c.cell == cell
  )
  if not choice_n:
```

```python
    choice_n = ChoiceNode(cell=cell)
    node.children.append(choice_n)

  sum_n = choice_n.children.get(letter)
  if not sum_n:
    sum_n = SumNode()
    choice_n.children[letter] = sum_n

  return add_word(sum_n, path[1:], points)
```

A Sum/Choice tree is an efficient way to represent the words in a Boggle board class because it stores each unique path to a word once, even though this same path may be valid for many individual boards in the class.

Next we define the "Force" operation, $F$. This sums all the paths to words found on a specific board class.

$$F(n : \text{SumNode}, B) := n.\text{points} + \sum_{n.\text{children}} F(c, B)$$

$$F(n : \text{ChoiceNode}, B) := \begin{cases} F(n.\text{children}[B_{n.\text{cell}}], B) \\ \quad \text{if } B_{n.\text{cell}} \in n.\text{children} \\ 0 \text{ else} \end{cases}$$

Intuitively, this "forces" each cell to match the board $B$, discarding all paths in the tree that don't match the board.

**Lemma**:

$$\text{F}(\text{add\_word}(T, P, \text{points}), B) =$$
$$\text{F}(T, B) + (\text{points if } P \in B \text{ else } 0)$$

Here $P \in B$ means that the path is compatible with the board, that is to say:

$$(P \in B) := B_{\text{cell}} = \text{letter} \ \forall (\text{cell}, \text{letter}) \in P$$

This result follows directly from the definitions of add_word and $F$. This lemma tells us that the Sum/Choice tree acts as a container structure for paths to words on a Boggle board.

Before building a tree for a board class, we make one more critical observation:

**Lemma**: $P \in B$ is independent of the order of $P$.

This follows immediately from the definition of $P \in B$, which is an "and" across the cells in the path. "And" is commutative, and hence compatibility is not dependent on the order of the cells in the path.

This means that, when we add a word to a tree, we're free to permute its cells in any way we like. In particular, we can establish a canonical order of the cells, so the same cells always appear at the top of the tree, as in Table 11.

| 3 | 7 | 5 | 2 |
|---|---|---|---|
| 11 | 15 | 13 | 10 |
| 9 | 14 | 12 | 8 |
| 1 | 6 | 4 | 0 |

Table 11: The canonical ORDER array for 4x4 Boggle. Higher numbered cells are sorted to the start of a path.

We're now ready to build trees.

```python
# Listing 4: Building a tree
def build_tree(board_class, trie):
  root = SumNode()

  def choice_step(idx, trie, choices):
    letters = board_class[idx]
    for letter in letters:
      if trie.has_child(letter):
        choices.append((idx, letter))
        child = trie.child(letter)
        sum_step(idx, child, choices)
        choices.pop()

  def sum_step(idx, trie, choices):
    if trie.is_word():
      ordered_choices = sorted(
        choices, key=lambda c: -ORDER[c[0]]
      )
      score = SCORES[trie.length()]
      add_word(root, ordered_choices, score)
    for n_idx in NEIGHBORS[idx]:
      if n_idx not in (c[0] for c in choices):
        choice_step(n_idx, trie, choices)

  for i in range(m * n):
    choice_step(i, trie, [])
  return root
```

Next, we can define an "upper bound" operation, $U$, on sum/choice trees:

$$U(n : \text{SumNode}) := n.\text{points} + \sum_{n.\text{children}} U(c)$$

$$U(n : \text{ChoiceNode}) := \max_{c \in n.\text{children}} U(c)$$

To show that this is a valid upper bound, we'll explore its relationship with the Force operation, $F$.

**Lemma**: $U(T) \geq F(T, B) \ \forall T, B$

For a SumNode, the definition of $U$ and $F$ are identical. For a ChoiceNode, $F$ picks an individual child, whereas $U$ takes the max across all its children. Hence $F(n) \leq U(n)$ for ChoiceNodes as well.

**Lemma**: $F(\text{BuildTree}(C), B)$ is independent of $C$

From the earlier lemma, $F(\text{BuildTree}(C), B)$ is a sum across paths to words found in $B$. The paths involving other possible letter choices are zeroed out. Hence $F(\text{BuildTree}(C), B)$ is a function of $B$ alone.

So what is $F(T, B)$? Looking at Listing 4, we can treat `choice_step` as picking $B_{\text{idx}}$, rather than iterating over the possible letters on a cell in the board class. Then the code is identical to `score` (Listing 0), except without the check for whether a word has been found before. It's finding the words on a Boggle board, except you're allowed to find the same word twice.

Hence we call this the "Multiboggle Score," $M(B)$:

$$M(B) := F(T, B)$$

**Lemma**: $M(B) \geq S(B)$

This is clearly true, since the regular Boggle score counts each unique word once, whereas the Multiboggle score might count it multiple times. If a board does not contain any repeat letters, then $M(B) = S(B)$.

**Theorem**: If $T = \text{BuildTree}(C)$, then

$$U(T) \geq S(B) \ \ \forall B \in C$$

That is to say, $U$ is a true upper bound. This follows from combining the previous lemmas:

$$\begin{aligned} U(T) &\geq F(T, B) \\ &= M(B) \\ &\geq S(B) \ \ \forall B \in C \end{aligned}$$

With these definitions and results, we can look at some of these trees and try to build an intuition for them.

Here's a small 2x2 board class containing two individual boards:

| T | I |
|---|---|
| {A, E} | R |

Figure 1 and Figure 2 show two trees for this board class, the first without word reordering ("spelling order") and the second with reordering (an "orderly tree").

We can make a few observations about these Sum/Choice trees:

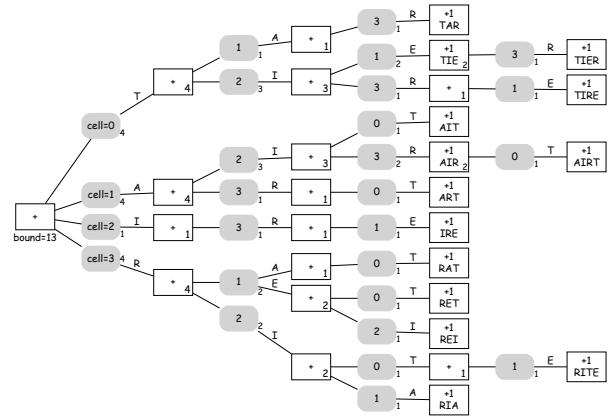

Figure 1: "Spelling order" tree for "t ae i r"; The bound `U(n)` is marked next to each node.

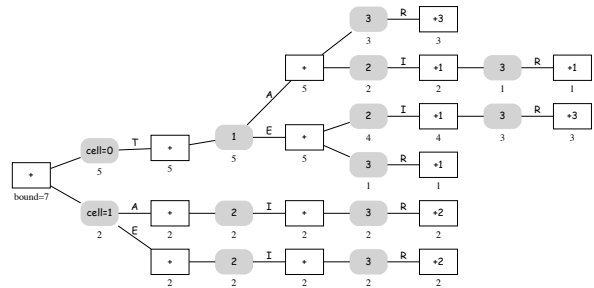

Figure 2: Orderly Tree for "t ae i r"; The bound `U(n)` is marked under each node.

- The wordlist and geometry of the Boggle board are fully encoded in the tree. Once the tree is constructed, we no longer need to reference the Trie or the `NEIGHBORS` array.
- In the "spelling order" tree, SumNodes with points correspond to individual words. Each path to a word is associated with a single SumNode.
- In the orderly tree, a SumNode may have multiple words associated with it. For example, the "+3" node on the top right of Figure 2 includes the words TAR, RAT and ART. If you can find one of these, you can find all of them.
- The orderly tree has significantly fewer nodes (29 vs. 56) and a lower bound (7 vs 13). It has two ChoiceNodes for cell 1, compared with 7 for the "spelling order" tree.
- The orderly tree for a board class, and hence the "orderly bound," is dependent on the canonical order that we choose for the cells.

Both of the boards in this board class score seven points, so the orderly tree's bound is "tight." This isn't always the case, however. ChoiceNodes for the same cell may appear multiple times in the tree. The bound may be imprecise because the max operation may not make the same choice on each ChoiceNode.

On this board class, for example:

| T | E |
|---|---|
| {A, O} | D |

- There are two words that use all four cells with an "O:" DOTE and TOED. But only DATE uses all four with an "A." So for this set of cells, it's better to pick "O" than "A."
- There are many words that use "T," "E" and "A:" ATE, ETA, EAT, TEA. But there's only one word that uses "T," "E" and "O:" TOE. So for these three cells, it's better to pick "A" than "O."

These two are summed. Of course, the cell can't be both an "A" and an "O" at the same time, so this inconsistent choice results in an overcount. The bound is 10, whereas both boards score 8. (The "spelling order" bound is 11.)

Ordering the paths helps for these small board class, but the effect is more dramatic for larger board classes:

**Bound on Root**

| Board | "Spelling" | "Orderly" |
|---|---|---|
| 2x2 | 13 | 7 |
| 3x3 (a) | 6,361 | 503 |
| 3x3 (b) | 9,460 | 1,523 |
| 3x4 (a) | 51,317 | 4,397 |
| 3x4 (b) | 194,425 | 10,018 |
| 3x4 (c) | 69,889 | 4,452 |
| 4x4 (a) | 176,937 | 11,576 |
| 4x4 (b) | 514,182 | 53,037 |

Table 14: Examples of bounds on the trees for various board classes. The bound on "orderly" trees is consistently much lower than that of the "spelling order" trees.

The bound might also be an overcount if there are repeat letters and we double-count a word. When we work with Sum/Choice trees, we are fundamentally working with the Multiboggle score, rather than the true Boggle score. For most boards, $M(B)$ is close to $S(B)$. Since we have considerable "wiggle room" between the average score of a board (~40 points) and the score of the best board (3625), working with the Multiboggle score is usually an acceptable concession. What we'll seek is boards $B$ with $M(B) \geq S_{\text{high}}$. For each of these, we can confirm whether $S(B) \geq S_{\text{high}}$ as well using a regular Boggle solver.

While $M(B)$ is usually close to $S(B)$, there are some pathological cases where this breaks down. For example, the board in Table 15 has $S(B) = 189$, but $M(B) = 21,953$! (The word "reservers" can be found in 100 distinct ways.) We will partially address this issue later in the paper.

| E | E | E | S |
|---|---|---|---|
| R | V | R | R |
| E | E | E | S |
| R | S | R | S |

Table 15: Pathological board with Score(B)=189 but Multi(B)=21,953.

Sum/Choice trees are reminiscent of Algebraic Decision Diagrams (ADDs),[11] particularly in the way that fixing an order for the decisions results in more compact, efficient trees.[12] The fundamental difference is the Sum nodes. To evaluate an ADD for a given set of choices, you descend the tree to a single, terminal node. Evaluating a Sum/Choice tree requires adding values from many subtrees.

ADDs are "reduced" by factoring out identical subtrees to produce a DAG. This could be done with Sum/Choice trees as well as a way to save memory, but this did not prove necessary to solve the 4x4 Boggle maximization problem.

Both Sum/Choice trees and ADDs are graphical representations of functions, in this case $M(B)$. Constructing an ADD for this function would be impractical, because it would require an exponential number of edges. Since we don't care about $M(B)$ for each $B$, only a bound on it, building a tree with each $M(B)$ in a terminal node would be wasteful. The Sum nodes result in a more compact representation at the cost of being more difficult to work with.

**Sum/Choice Satisfiability is NP-Hard**

We seek boards $B$ in a board class $C$ such that $F(T, B) \geq S_{\text{high}}$. Since each board in a board class represents a choice of letter on each of the cells, we can think of this as a satisfiability problem.

**Theorem**: Determining whether there exists $B$ such that $F(T, B) \geq S_{\text{high}}$ is NP-Hard.

Proof: We map from 3-CNF, a known NP-Hard problem, to the Sum/Choice Tree satisfiability problem.

Suppose we have a 3-CNF formula with $m$ clauses on $x_1, x_2, ..., x_n$.

For each clause, we construct a tree which evaluates to 1 if the clause is satisfied and zero if it is not satisfied.

- If the clause contains a single term $a$, then we model this as a ChoiceNode on cell $a$.
- If the clause is $a \vee b$, we model is as a tree with two layers of ChoiceNodes.
- If the clause is $a \vee b \vee c$, we model it as a tree with three layers of ChoiceNodes.

Finally, we create a root SumNode $T$ with the $m$ ChoiceNodes as children. By construction, $\exists B \mid F(T, B) = m$ iff there are $x_i$ that satisfy the 3-SAT problem. So if we can solve the satisfiability problem for Sum/Choice trees, we can also solve it for 3-CNF. Since 3-CNF is known to be NP-Hard, this means that Sum/Choice satisfiability is NP-Hard as well.[1]

So we should not expect to find an efficient solution to this problem, nor one that scales well to larger boards. This doesn't necessarily mean that Boggle maximization itself is NP-Hard, since not every Sum/Choice tree corresponds to a Boggle board. Still, it is suggestive that this is a hard problem.

**Merge Operation**

With Sum/Choice Trees defined, we're ready to perform operations on them. Our first goal will be to speed up the "branch" operation and calculation of the subsequent bound. This requires forcing a single cell in the board class to be each of its possible letters and constructing the resulting trees.

In practice, it makes the most sense to force the top choice in the tree, i.e. the one with the first position in the canonical order. This requires a "merge" operation on orderly trees, which is straightforward to implement as in Listing 5.

```python
# Listing 5: merge operation on orderly trees
def merge(a: SumNode, b: SumNode) -> SumNode:
  by_cell = {c.cell: c for c in a.children}
  for bc in b.children:
    ac = by_cell.get(bc.cell)
    by_cell[bc.cell] = (
      merge_choice(ac, bc) if ac else bc
    )
  ch = [*by_cell.values()]
  return SumNode(
    points=a.points + b.points, children=ch
  )

def merge_choice(
  a: ChoiceNode, b: ChoiceNode
) -> ChoiceNode:
  ch = {**a.children}
  for choice, bc in b.children.items():
    ac = ch.get(choice)
    ch[choice] = merge(ac, bc) if ac else bc
  return ChoiceNode(cell=a.cell, children=ch)
```

**Lemma**: $\forall B \in T_1, T_2$

$$F(\text{merge}(T_1, T_2), B) = F(T_1, B) + F(T_2, B)$$

No cells are destroyed by merge, and points are added when there's a collision.

With the merge helper, we can define the "branch" operation. Note that N here isn't really a free parameter; if we have an orderly tree with $N$ cells, then we must use this value of $N$.

```python
# Listing 6: branch operation on orderly trees
def branch(
  o: SumNode,
  N: int,
  board_class: list[str],
) -> list[SumNode]:
  top_choice = find(
    o.children, lambda c: c.cell == N
  )
  if not top_choice:
    # cell is irrelevant;
    return [o for _ in board_class[N]]

  other_choices = [
    c for c in o.children if c.cell != N
  ]
  skip_tree = SumNode(
    children=other_choices, points=o.points
  )
  return [
    merge(
      top_choice.children[letter], skip_tree
    )
    if top_choice.children.get(letter)
    else skip_tree  # dead letter on cell N.
    for letter in board_class[N]
  ]
```

---

[1]This proof was provided by D.W. on StackOverflow, see [13]

```
    ]
```

The `branch` function splits the tree into two parts: one that includes the words that go through the "top" cell (`top_choice`) and another (`skip_tree`) that includes all the words that don't. Each child of `top_choice` corresponds to a particular choice of letter on that cell. Every word must fall into one of these two groups (goes through the cell or doesn't). Adding their bounds will produce a valid bound for this choice of letters. Since the `merge` operation distributes over the Force function $F$, the resulting tree will have a valid bound.

Calling `branch` is considerably faster than building a new tree for each letter choice on a cell. For example, on the high-scoring 4x4 board class from Table 14, `branch` split a center cell containing 12 letters and returned 12 subtrees in 0.07s. Building the same trees from scratch took 4.0s, roughly a 60x difference.

A few observations:

- The `merge` operations perform a deep merge.
- In practice we can update a `bound` property on all nodes as we merge them, so that calculating the bound is instant.
- The `merge` operation (and `branch`) are likely to reduce the bound because they synchronize choices across previously distinct subtrees.
- The subtrees we get back from `branch` aren't exactly the same as what you'd get by building orderly trees for the smaller board classes directly. This is because `branch` effectively removes a cell from the board, rather than replacing it with a one-letter choice.
- The `branch` operation can be thought of as "pivoting" a choice to the top of the tree, making it more like a true decision diagram.

Here are the results of the first `branch` call on the tree for the 5 million board 3x3 class from Table 6.

| Center Cell | Nodes | Bound | True Max |
|---|---|---|---|
| aeiou | 333,492 | 1,523 | 545 |
| a | 86,420 | 1,198 | 545 |
| e | 98,585 | 1,417 | 520 |
| i | 81,062 | 994 | 503 |
| o | 75,474 | 862 | 392 |
| u | 60,457 | 753 | 326 |

**Orderly Bound Operation**

The `branch` operation on its own is sufficient to implement Branch and Bound with Sum/Choice Trees. Eventually we'll merge all the way down to a single SumNode, which represents a candidate board. This allocates additional nodes, however, and we may wish to save RAM and compute by not doing that all the way to the bottom.

Instead, we can define an alternative algorithm, `orderly_bound`, which refines the bound on a tree by traversing it, rather than merging subtrees. We'll maintain a stack of pointers to ChoiceNodes. To "branch,"" we'll pop off the ChoiceNodes for the next cell and, for each choice, push all the next child cells. We can maintain a bound as we do this. If the bound ever drops below $S_{\text{high}}$, we can abandon this search path.

```
# Listing 7: orderly_bound
# Assumes N >= 1
def orderly_bound(
  root: SumNode,  # Orderly(N)
  board_class: list[str],
  S_high: int,
):
  def step(
    points: int,
    idx: int,
    # letters chosen on previous cells
    choices: list[char],
    stack: list[ChoiceNode],
  ):
    b = points + sum(bound(n) for n in stack)
    if b < S_high:
      return  # This path is eliminated
    if idx == N:
      # complete board that can't be eliminated
      record_candidate_board(choices, b)
      return

    # Try each letter on the next cell in order.
    cell = CELL_ORDER[idx]
    for letter in board_class[cell]:
      next_nodes = [
        n for n in stack if n.cell == cell
      ]
      next_stack = [
        n for n in stack if n.cell != cell
      ]
      next_points = points
      next_choices = choices + [letter]
      for node in next_nodes:
        letter_node = node.children.get(letter)
        if letter_node:
          next_stack += letter_node.children
          next_points += letter_node.points
```

```
    step(
      next_points,
      idx + 1,
      next_choices,
      next_stack,
    )

  step(root.points, 0, [], root.children)
```

**Lemma**: Each `step` call preserves the invariant that

$$\text{points} + \sum_{n \in \text{stack}} F(n, B) = M(B)$$

for all boards $B$ compatible with `choices`.

The proof is by induction, and is omitted for brevity.

**Theorem**: `orderly_bound` finds all the boards $B$ in a tree with $M(B) \geq S_{\text{high}}$.

Proof: The lemma established an invariant for the recursive calls to `step`. It suffices to check the check the two cases where the function returns early.

If $b < S_{\text{high}}$, then we have:

$$S_{\text{high}} > b = \text{points} + \sum_{c \in \text{stack}} U(c)$$
$$\geq \text{points} + \sum_{c \in \text{stack}} F(c, B)$$
$$= M(B) \ \forall B$$

and therefore there are no high-scoring boards.

If `idx==N`, then the stacks are empty and we have a single board with

$$b = \text{points} + \sum_{c \in \text{stack}} U(c)$$
$$= \text{points}$$
$$= M(B) \geq S_{\text{high}}$$

So this is a candidate high-scoring board.

Observations:

- `orderly_bound` performs a shallow merge.
- `orderly_bound` stores at most one pointer to each node in the tree, but in practice many fewer. For the board class from Table 6, the maximum stack size is 167.
- The `points` parameter to `step` is the Multiboggle score on the portion of the board that's been forced.
- `orderly_bound` has exponential backtracking behavior. We visit a node in `stack` something like 2^(# of skipped nodes). For the board class from Table 6, the most times a single node is visited is 1,289.
- The `branch` operation (tree merging) mitigates this exponential by reducing the number of skips.
- We used a single stack here, but in practice it's much more efficient to keep a separate stack for each cell and cache the sums.

The `branch` and `orderly_bound` operations work well together. In practice, we build the tree for a board class, then call `branch` some number of times before switching over to `orderly_bound`. The optimal switchover point is highly variable, but switching when $n.\text{bound} \leq 1.5 S_{\text{high}}$ or there are only four unmerged cells left seems to work well in practice (see Table 17).

| Switchover Score | Time (s) | RAM (G) | Nodes Allocated |
|---|---|---|---|
| $\infty$ (Bound) | 3629 | 1.65 | 89,638,760 |
| 10,000 | 152.8 | 2.08 | 269,306,267 |
| 7,500 | 76.93 | 2.09 | 452,922,484 |
| 6,000 | 57.76 | 2.07 | 849,594,151 |
| 5,000 | 65.49 | 2.07 | 1,554,322,820 |
| 4,000 | 114.0 | 2.10 | 3,403,803,368 |
| 0 (Branch) | 175.4 | 2.06 | 5,406,460,230 |

Table 17: The effect of switchover score on runtime and memory usage for a single board class. $S_{\text{high}}$ was 3,500. The initial bound was 30,806. Nodes is the total number of nodes that were allocated during Branch and Bound, not peak nodes.

**De-duplicated Multiboggle**

`orderly_bound` will report any board with $M(B) \geq S_{\text{high}}$. For each of these, we need to check whether $S(B) \geq S_{\text{high}}$ as well. As we saw earlier, these two scores are typically close, but there are some pathological cases where they diverge. Since `branch` and `orderly_bound` converge on the Multiboggle score, they'll bog down on the board classes containing these highly duplicative boards, and we'll wind up having to score many millions of Boggle boards to filter them out.

We can improve the situation slightly. Here's a 2x3 board containing three words, BEE, FEE, and BEEF.

| E | B | E |
|---|---|---|
| E | F | E |

The Multiboggle score counts each of these words four times. If we consider this in the context of a board class, however:

| {E,X} | B | {E,X} |
|---|---|---|
| {E,X} | F | {E,X} |

we can see that BEE can be found on the left only when both cells are E, not X, and similarly on the right. But when both of the left cells are E, we know that these paths to BEE, FEE and BEEF will only count once towards the true Boggle score. Both of the left paths to BEE:

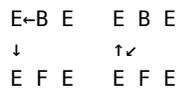

correspond to the exact same SumNode in the orderly tree. So if we add this word's points to that node once, rather than twice, we don't risk compromising our upper bound.

We do still need to add both the left and the right versions of BEE, FEE and BEEF. If we only added the left versions, then we'd miss the points for this board in the board class:

| X | B | E |
|---|---|---|
| X | F | E |

And we'd no longer have a valid upper bound.

We can call this revised Multiboggle score $D(B)$ (for "de-duped"). To calculate it, we only score words once for each unique (unordered) set of cells on which they can be formed. So for this board we have:

- $S(B)$ = 3
- $M(B)$ = 12
- $D(B)$ = 6

Clearly $S(B) \leq D(B) \leq M(B)$ for all boards B. For boards without repeated letters, $S(B) = M(B)$, and so the same holds for $D(B)$.

We can filter out duplicate words in `BuildTree`. All the same invariants now hold, only we converge to $D(B)$ rather than $M(B)$.

Here are some examples of the effect this deduping has on the root bound for Orderly Trees:

- eeesrvrreeesrsrs (Table 15): 21,953 → 13,253 (true score is 189)
- best 4x4 board class: 53,037 → 36,881 (max is 3,625)

Since the bound for this board class is so much lower, we expect the Branch and Bound procedure to process it much more quickly. In practice, this is a 5x speedup on this board class. This optimization has the greatest impact on the highest-scoring board classes, which take the most time to process.

**Final Branch and Bound Algorithm**

This is the final Branch and Bound algorithm for finding Boggle boards $B$ with $S(B) \geq S_{\text{high}}$:

1. Enumerate all possible board classes, filtering for symmetry.
2. For each board class $C$, build a Sum/Choice tree with deduping.
3. Repeatedly call `branch` until either:
   1. $U(\text{node}) < S_{\text{high}}$ in which case this board class can be eliminated.
   2. $U(\text{node}) \leq 1.5 S_{\text{high}}$ in which case we switch to `orderly_bound`. This will output a list of boards $B \in C$ such that $D(B) \geq S_{\text{high}}$.
4. For each such board $B$, check whether $S(B) \geq S_{\text{high}}$.

This will produce a list of all boards $B$ (up to symmetry) with $S(B) \geq S_{\text{high}}$. If two congruent boards fall in the same board class, it will produce both of them.

In practice, the individual board classes can be treated as independent tasks in a MapReduce.

## Results

The implementation can be found at https://github.com/danvk/hybrid-boggle/. The code is a mixture of C++ and Python, with the glue provided by pybind11. [14] Smaller runs were performed on the author's M2 MacBook Air, while larger runs were performed on C4 VMs (Intel Xeon) on Google Cloud Platform.

**Results for 3x3 and 3x4 Boggle**

The Branch and Bound procedure based on orderly trees is dramatically faster than brute force search. For 3x3 Boggle with two letter buckets on a single core, the Branch and Bound procedure completes in just 70 seconds. Compared to the 12 days it would have taken for exhaustive search, this represents a 15,000x speedup.

Using ENABLE2K, this search finds 42 distinct boards (up to symmetry) that score 500 points or more. Each of these boards can also be found via the hillclimbing procedure (Listing 1).

For 3x4 Boggle, we use two letter buckets in the four corners and three buckets for the other eight cells. The Branch and Bound procedure completes in 5h54m on a single core. This represents a 3,000,000x speedup compared to the 2,000 CPU years that exhaustive search would have required.

This search finds 33 distinct boards that score 1,500 points or more. Again, each of these boards can also be found via the hillclimbing procedure. This gives us confidence that it is an effective way to find the global maximum.

**Results for 4x4 Boggle**

Three 4x4 runs were completed, with the ENABLE2K, OSPD5 and NASPA2023 word lists. The same bucketing was used as 3x4: two buckets in the corners, three in the other cells. This results in just over one million canonically-oriented board classes.

- **ENABLE2K**: Found 32 boards with $S(B) \geq 3500$ in 23,000 CPU hours[2]
- **OSPD5**: Found 46 boards with $S(B) \geq 3600$ in 7,500 CPU hours
- **NASPA2023**: Found 40 boards with $S(B) \geq 3700$ in 9,000 CPU hours.

Compared to exhaustive search, this is roughly a billion times faster. Assuming $0.05/core/hr, this is around $400 of compute in 2025.

As with 3x3 and 3x4 Boggle, the top boards can all be found via the hillclimbing procedure in Listing 1.

Here are the top five boards for ENABLE2K and NASPA2023:

|   | ENABLE2K | NASPA2023 |
|---|---|---|
| 1 | perslatgsineters: 3625 | perslatgsineters: 3923 |
| 2 | segsrntreiaeslps: 3603 | segsrntreiaeslps: 3861 |
| 3 | gepsnaletireseds: 3593 | bestlatepirsseng: 3841 |
| 4 | aresstapenildres: 3591 | dclpeiaerntrsegs: 3835 |
| 5 | cinslateperidsng: 3591 | aresstapenildres: 3826 |

There is considerable overlap between the highest-scoring boards for each wordlist. ENABLE2K and NASPA2023 share a top board. The top board for OSPD5 is the #2 board for ENABLE2K.

---

[2]This run predated the deduped multiboggle optimization, so it ran considerably slower than the other runs.

**Extension to maximizing word count**

Instead of seeking the highest-scoring board, we might instead be interested in finding the board with the most words on it. This is a straightforward modification of the problem. We simply change the SCORES array to contain all 1s. Then all the tools developed in this paper work exactly as before.

Hillclimbing is also effective at solving this problem. For 3x4 Boggle, the "wordiest" board found via hillclimbing matches the global max found via Branch and Bound. While no exhaustive search has been conducted for the wordiest 4x4 board, we'd expect the hillclimbing winner to be the global optimum here as well.

These boards have significant overlap with the highest-scoring boards. Table 22 shows the wordiest board for ENABLE2K. This is also the #8 highest-scoring board. Its middle two columns are identical to those of the highest-scoring board, which contains 1,045 words.

| S | E | R | G |
|---|---|---|---|
| L | A | N | E |
| P | I | T | S |
| S | E | R | O |

Table 22: The "wordiest" known board for ENABLE2K with 1,158 words and 3,569 points.

**Variation: Powers of Two Boggle**

It's an interesting quirk of Boggle that you score more points for longer words, but only up to eight letters. What if you kept getting more points for longer words? We can set SCORES=[0, 0, 0, 1, 2, 4, 8, 16, 32, …] to play "powers of two" Boggle, where a three letter word is still worth one point, but an eight letter word is worth 32, a sixteen letter word is worth 8,192 points and a seventeen letter word (which must contain a Qu) is worth 16,384 points.

Hillclimbing is *not* effective at finding the highest-scoring board in this version of Boggle. The best board found in 50 hillclimbing runs was cineqetnsniasesl (28,542 points). However, we can exhaustively search all boards containing a 17-letter word to find rpqaselinifcoita (44,726 points, Table 23). Over a third of this board's points come from the single 17-letter word "prequalifications."

| R | P | Qu | A |
|---|---|----|---|
| S | E | L  | I |
| N | I | F  | C |
| O | I | T  | A |

Table 23: The best known board for "powers of two" Boggle, containing the 16,384-point word "prequalifications."

This failure gives us some insights into why hillclimbing is effective at finding the highest-scoring and wordiest boards. Those score functions both produce a smooth fitness landscape, where single character variations on a board produce relatively small changes to its score. This means that the optimal boards are surrounded by other high-scoring boards, and hillclimbing is likely to reach the summit if it gets anywhere near it. There's nowhere for a great board to "hide," surrounded by mediocre boards.

Contrast this with powers of two Boggle, where a one character change is likely to remove the longest word on a great board, greatly reducing its score.

**Future Work**

Branch and Bound with Sum/Choice trees is extremely effective at exploiting the structure inherent in Boggle, making global searches over the 4x4 board space feasible in a data center. This still takes a lot of compute, however, enough that full searches have not been conducted for every wordlist, for other languages, or for the most word-dense boards. Further incremental optimizations and reductions in compute costs might make this more attractive.

There is also a 5x5 version of Boggle ("Big Boggle"), and there was a limited run of a 6x6 version ("Super Big Boggle"). These are much harder problems. Even with all the optimizations presented in this paper, an exhaustive search for the best 5x5 Boggle board is expected to take on the order of a billion CPU years. This suggests that a radically different approach is required.

| Size | CPU time |
|------|----------|
| 3x3  | 70s      |
| 3x4  | 6h       |
| 4x4  | 9,000h   |
| 4x5  | ~10,000 years |
| 5x5  | ~1B years |

Table 24: Branch and Bound search time for various board sizes. Times for 4x5 and 5x5 are estimates.

Here are a few alternative approaches that would be interesting to explore:

- **SMT/ILP solvers** The NP-Hard proof mapped Boggle maximization onto well-known SAT problems. These problems are often tackled using ILP Solvers like Z3, OR-Tools, cvc5 or Gurobi. A preliminary investigation didn't show much promise here[15], but this is a large field and more experienced practitioners may be able to use these tools to greater effect.
- **GPU acceleration** The process described in this paper relies entirely on the CPU. But the biggest advances in recent years have come from GPUs. It's not immediately clear how Boggle maximization as described here could be GPU accelerated since the algorithm is tree-heavy and full of data dependencies. Still, there might be another way to frame the problem that results in better acceleration.

| L | I | G | D | R |
|---|---|---|---|---|
| M | A | N | E | S |
| I | E | T | I | L |
| D | S | R | A | C |
| S | E | P | E | S |

Table 25: Best known 5x5 board for ENABLE2K, with 2,344 words and 10,406 points[1]

**Conclusions**

Using Branch and Bound, Sum/Choice trees, and some tailor-made algorithms, we're able to achieve a factor of a billion speedup over brute-force search. This is enough to prove that the highest-scoring board found via hillclimbing is, in fact, the global optimum. It's likely that hillclimbing is so effective because the score function produces a relatively smooth fitness landscape. The approach taken in this paper requires solving an NP-Hard problem, and it will not scale well to 5x5 or 6x6 Boggle maximization, which remain well out of reach.